\documentclass{iopart}
\usepackage{graphicx,epsfig}
\usepackage{bm}
\usepackage{iopams}
\newcommand{\ba}{\begin{eqnarray}}
\newcommand{\ea}{\end{eqnarray}}
\newcommand{\bse}{\numparts}
\newcommand{\ese}{\endnumparts}

\newcommand{\DD}{{\cal {D}}}

\newcommand{\bbq}{\begin{quote}}
\newcommand{\eeq}{\end{quote}}
\newcommand{\RR}{{}^3{\cal{R}}}

\newcommand{\T}{{}^3{\cal{T}}}

\newcommand{\EE}{{\cal{E}}}

\newcommand{\JJ}{{\cal{J}}}

\newcommand{\HH}{{\cal{H}}}

\newcommand{\PP}{{\cal{P}}}

\newcommand{\Da}{\delta^{(A)}}

\newcommand{\Dth}{\delta^{(\Theta)}}
\newcommand{\dDth}{\dot\delta^{(\theta)}}
\newcommand{\Dm}{\delta^{(\mu)}}
\newcommand{\dDm}{\dot\delta^{(\mu)}}
\newcommand{\dDn}{\dot\delta^{(n)}}

\newcommand{\Dp}{\delta^{(p)}}
\newcommand{\Dn}{\delta^{(n)}}
\newcommand{\dDp}{\dot\delta^{(p)}}

\newcommand{\DRR}{\delta^{(\RR)}}
\newcommand{\dd}{{\rm{d}}}

\begin{document}


\title[Shear viscosity, relaxation and collision times in spherically symmetric spacetimes.]{Shear viscosity, relaxation and collision times in spherically symmetric spacetimes.} 
\author{ Roberto A. Sussman}
\address{Instituto de Ciencias Nucleares, Universidad Nacional Aut\'onoma de M\'exico (ICN-UNAM),
A. P. 70--543, 04510 M\'exico D. F., M\'exico. }
\ead{sussman@nucleares.unam.mx}
\date{\today}
\begin{abstract}  We interpret as shear viscosity the anisotropic pressure that emerges in inhomogeneous spherically symmetric spacetimes described by the Lema\^\i tre--Tolman--Bondi (LTB) metric in a comoving frame. By assuming that local isotropic pressure and energy density satisfy a generic ideal gas equation of state, we reduce the field equations to a set of evolution equations based on auxiliary quasi--local variables. We examine the transport equation of shear viscosity from Extended Irreversible Thermodynamics and use a numerical solution of the evolution equations to obtain the relaxation times for the full and ``truncated'' versions. Considering a gas of cold dark matter WIMPS after its decoupling from the cosmic fluid, we show that the relaxation times for the general equation are qualitatively analogous to collision times, while the truncated version is inadequate to describe transient phenomena of transition to equilibrium.                
\end{abstract}

\section{Introduction.}
 
It is a well known fact that dissipative effects in the context of General Relativity must comply with causality and stability requirements~\cite{rund,JCVL,fundam1,fundam2}. Also, there is an evident theoretical connection between dissipative phenomena and anisotropy or inhomogeneity of self--gravitating sources. This emerges from the fact that heat flux and shear viscosity couple with the 4--acceleration, shear and spacelike gradients in their corresponding  evolution (or transport) equations. Since bulk viscosity is the only dissipative stress compatible with global isotropy and homogeneity,  most articles on dissipative cosmological sources deal with the effects of this stress in a Friedman--Lema\^\i tre--Robertson--Walker (FLRW) context \cite{FLRW}. However, the literature contains also a large number of studies of dissipative cosmological sources under anisotropic and inhomogeneous conditions, using Bianchi or Kantowski--Sachs models~\cite{anisinhom1}, involving heat flux~\cite{anisinhom2} or shear viscosity with the Lemaitre--Tolman--Bondi metric~\cite{anisinhom3,anisinhom4,anisinhom5} (see also \cite{kras} for inhomogeneous spacetimes with dissipative sources).     

Besides mathematical simplicity, the main justification for preferring a FLRW framework, or linear perturbations on a FLRW background, in cosmological studies is the conjecture (supported by observations) that the universe is approximately FLRW at a large ``homogeneity'' scale of 150-300 Mpc~\cite{review}. Thermal dissipation might play a minor role in these scales, as observations seem to reveal that cosmic dynamics is presently dominated by (apparently) non--thermal sources (cold dark matter and dark energy \cite{review}).  While dissipative phenomena of a thermal nature are relevant for understanding early universe interactions, inhomogeneity can be safely assumed to be very small in these conditions. Dissipative phenomena also arise in self--gravitating (and inhomogeneous) sources at local scales, either stellar or galactic~\cite{rund}. In some cases (inter--stelar or inter--galactic clouds of ionized gas), characteristic velocities and energies are basically non--relativistic, but in other cases (gas accretion to compact objects or AGN's, jets, photon or neutrino transport) we can have non--trivial relativistic and ultra-relativistic effects in conditions of non--linear inhomogeneity~\cite{padma2}. However, as long as we ignore the fundamental physics of dark matter and dark energy, we can still try to probe theoretically the possibility of some forms of thermal dissipation in these sources and/or their interactions at the cosmic scale.          
  
Fully general inhomogeneity requires numerical codes of high complexity, hence we offer in this article a compromise by looking at dissipative phenomena in spherically symmetric sources, which in spite of their obviously idealized nature, are still useful to examine non--linear phenomena that cannot be studied in a FLRW framework or with linear perturbations. By considering ``LTB spacetimes'' that generalize to nonzero pressure the well known LTB dust solutions~\cite{sussQL}, we obtain a class of spacetimes that can be fully described by autonomous first order evolution equations that can be well handled by simple numerical methods. These models are quite general and readily allow for an inhomogeneous generalization of a large number of known FLRW solutions. The reader can consult \cite{sussQL} for an extended and comprehensive discussion on  these spacetimes and their physical and geometric properties. 

The plan of the article is as follows. We describe in section 2 the basic features of LTB spacetimes~\cite{kras, sussQL}, in which the anisotropic pressure is considered as a shear viscous stress~\cite{anisinhom3,anisinhom4,anisinhom5}. Assuming a conserved particle current and the entropy current associated with Extended Irreversible Thermodynamics, we derive in section 3 the full causal transport equation for shear viscosity~\cite{rund,JCVL,fundam1,fundam2}. In section 4, we provide the ``fluid flow'' evolution equations for LTB spacetimes~\cite{1plus3}, equivalent to the field equations, in terms of suitably defined quasi--local variables~\cite{sussQL}. In this description, the local thermodynamical state variables are gauge invariant ``exact'' perturbations of their quasi--local equivalents. In section 5 we specialize the evolution equations for a local equation of state corresponding to a generic ideal gas that covers the cases of (i) a classical ideal gas and (ii) the coupled mixture of a non--relativistic gas and radiation (the ``radiative gas'' \cite{rund,fundam1,anisinhom4,anisinhom5}). In section 6 we specialize the evolution and transport equations for the ideal gas, as a model of a gas of non--relativistic WIMPS after their decoupling from the cosmic fluid~\cite{rund,padma2,KT}, when particle numbers are conserved. We evaluate the relaxation times for the full transport equation and for its ``truncated'' version (the Maxwell--Cattaneo equation). In section 7 we compare numerically these times with mean collision times, showing that they are qualitatively analogous in the relaxation time scale. These numerical examples also show that the truncated equation is inadequate to describe the transient phenomena of transition to equilibrium for gas of WIMPS. This result is analogous to that obtained for the decoupling of matter and radiation in the radiative gas~\cite{anisinhom5}. We summarize the results obtained in section 8.        

\section{LTB spacetimes in the ``fluid flow'' description.}

Spherically symmetric inhomogeneous dust sources are usually described by the well known Lema\^\i tre--Tolman--Bondi metric~\cite{kras,sussQL}
\begin{equation}\dd s^2 = -c^2\dd t^2+ \frac{R'{}^2}{1-K}\dd r^2+R^2\left(\dd\theta^2+\sin^2\theta\dd\phi^2\right).\label{ltb}\end{equation}
where $R=R(ct,r)$,\, $R'=\partial R/\partial r$ and $K=K(r)$. A large class of spherically symmetric spacetimes follow at once by considering the most general source for (\ref{ltb}) in a comoving frame ($u^a=\delta^a_0$), which is the energy--momentum tensor
\begin{equation}T^{ab}=\mu\,u^au^b+p\,h^{ab}+\Pi^{ab}, \label{Tab}\end{equation}
where $\mu$ and $p$ are the matter--energy density and the isotropic pressure, $h^{ab}=u^au^b+g^{ab}$ is the induced metric of hypersurfaces $\T$ orthogonal to $u^a$, and $\Pi^{ab}$ is the symmetric traceless tensor of anisotropic pressure. We will call ``LTB' spacetimes'' to all solutions of Einstein's equations for (\ref{ltb}) and (\ref{Tab}). 

Besides the scalars $\mu$ and $p$, and the tensor $\Pi^{ab}$, the remaining basic covariant objects of LTB spacetimes are: 
\ba \Theta &=& \tilde\nabla_au^a=\frac{2\dot R}{R}+\frac{\dot R'}{R'},\qquad \hbox{Expansion scalar},\label{Theta}\\
\RR &=& \frac{2(KR)'}{R^2R'},\hskip 2.2cm \hbox{ Ricci scalar of the}\,\, \T,\label{RR}\\
\sigma_{ab} &=& \tilde\nabla_{(a}u_{b)}-\frac{\Theta}{3}h_{ab},\qquad\quad \hbox{ Shear tensor},\label{shear}\\
E^{ab} &=& u_cu_d C^{abcd},\hskip 2.2cm  \hbox{Electric Weyl tensor},\label{EWeyl}\ea
where $\dot R=u^a\nabla_a R$,\, $\tilde\nabla_a = h_a^b\nabla_b$,\, and $C^{abcd}$ is the Weyl tensor. 

For spherically symmetric spacetimes, the symmetric traceless tensors $\sigma^{ab},\,\Pi^{ab}$ and $E^{ab}$ can be expressed in terms of  single scalar functions as
\begin{equation} \sigma^{ab}=\Sigma\,\Xi^{ab},\qquad \Pi^{ab}=\PP\,\Xi^{ab},\qquad E^{ab}=\EE\,\Xi^{ab},  \label{PSEsc}\end{equation}
where $\Xi^{ab}=h^{ab}-3\eta^a\eta^b$ and $\eta^a=\sqrt{h^{rr}}\delta^a_r$ is the unit vector orthogonal to $u^a$ and to the 2--spheres orbits of SO(3) parametrized by $(\theta,\phi)$. 

The field equations  $G^{ab}=\kappa T^{ab}$ (with $\kappa=8\pi G/c^4$) for (\ref{ltb}) and (\ref{Tab}) are
\bse\ba \kappa\,\mu\,R^2R' &=& \left[R(\dot R^2+K)\right]',\label{mu1}\\
\kappa\,p\,R^2R' &=& -\frac{1}{3}\left[R(\dot R^2+K)+2R^2\ddot R\right]',\label{p1}\\
\kappa\,\PP\,\frac{R'}{R} &=& -\frac{1}{6}\left[\frac{\dot R^2+K}{R^2} +\frac{2\ddot
Y}{Y}\right]',\label{PP1}\ea\ese
From (\ref{Theta}), (\ref{mu1})--(\ref{PP1}) and (\ref{PSEsc}) we obtain the expressions for $\EE$ and $\Sigma$ in terms of metric functions
\begin{equation}\Sigma = \frac{1}{3}\left[\frac{\dot R}{R}-\frac{\dot R'}{R'}\right],\qquad
\EE = -\frac{\kappa}{2}\,\PP-\frac{\kappa}{6}\,\mu+ \frac{\dot
R^2+K}{2R^2}.\label{SigEE1}\end{equation}
The energy--momentum balance equations $\nabla_b T^{ab}=0$ for (\ref{Tab}) are
\bse\ba \dot\mu &=& -(\mu+p)\,\Theta-\sigma_{ab}\Pi^{ab}=-(\mu+p)\,\Theta-6\,\Sigma\,\PP,\label{ecbal1}\\
\tilde\nabla_a p &=& -\tilde\nabla_b\Pi^b\,_a,\qquad \Rightarrow\quad p'-2\,\PP' =6\,\PP\,\frac{R'}{R},\label{ecbal2}\ea\ese
so that pressure gradients are effectively supported by the anisotropic pressure.
 
Bearing in mind (\ref{PSEsc}) and the remaining previous equations, all covariant objects (scalars and proper tensors) in LTB spacetimes can be fully characterized by the following set of local covariant scalars:
\begin{equation}\{\mu,\,p,\,\PP,\,\Theta,\,\Sigma,\,\EE,\,\RR\}.\label{loc_scals}\end{equation} 
Given the covariant ``1+3'' time slicing afforded by $u^a$,  the evolution of these scalars can be completely determined by the following set of ``fluid flow''  scalar evolution equations~\cite{1plus3}
\bse\ba
\dot\Theta &=&-\frac{\Theta^2}{3}
-\frac{\kappa}{2}\left(\,\mu+3p\,\right)-6\,\Sigma^2,\label{ev_theta_13}\\
\dot \mu &=& -\,(\mu+p)\,\Theta-6\,\Sigma\, \PP,\label{ev_mu_13}\\
\dot\Sigma &=& -\frac{2\Theta}{3}\,\Sigma+\Sigma^2-\EE+\frac{\kappa}{2}\PP,
\label{ev_Sigma_13}\\
 \dot\EE &=& -\frac{\kappa}{2}\dot \PP-\frac{\kappa}{2}\left(\mu+p-2\PP\right)\,\Sigma
-3\left(\EE+\frac{\kappa}{6}\PP\right)\left(\frac{\Theta}{3}+\Sigma\right),\nonumber\\\label{ev_EE_13}\ea\ese
together with the spacelike constraints  
\bse\ba (p-2\PP)\,'-6\,\PP\,\frac{R'}{R}=0,\label{cPP_13}\\
\left(\Sigma+\frac{\Theta}{3}\right)'+3\,\Sigma\,\frac{R'}{R}=0,\label{cSigma_13}\\
\frac{\kappa}{6}\left(\mu+\frac{3}{2}\PP\right)'
+\EE\,'+3\,\EE\,\frac{R'}{R}=0,\label{cEE_13}\ea\ese
and the Friedman equation (or ``Hamiltonian'' constraint)
\begin{equation}\left(\frac{\Theta}{3}\right)^2 = \frac{\kappa}{3}\, \mu
-\frac{\RR}{6}+\Sigma^2,\label{cHam_13}\end{equation}
The system
(\ref{ev_theta_13})--(\ref{cHam_13}) is equivalent to the field plus conservation equations $\nabla_bT^{ab}=0$ (equations (\ref{ev_mu_13}) and (\ref{cPP_13})). However, this system requires an equation of state linking $\mu,\,p$ and $\PP$ to become determined, and the time and radial derivatives (in general) do not decouple. Hence, we will consider in section 5 another set of equivalent (but easier to handle) scalar evolution equations.

\section{Extended Irreversible Thermodynamics.}

In order to arrive to a determined set of evolution equations, we need to prescribe an equation of state that is suitable for a given physical model. If the desired model is a thermal system, it is necessary to consider $\mu$ and $p$ as thermodynamical scalars. In particular, a very useful system is the ideal gas associated with the following equilibrium equation of state~\cite{rund,anisinhom3}
\begin{equation}\mu = mc^2\,n+\frac{p}{\gamma-1},\qquad k\,T=\frac{p}{n} \label{eq_state}\end{equation}
where $n$ is the particle number density for a gas whose particles have mass $m$, $T$ is the temperature, $k$ is Boltzmann's constant and $\gamma$ is a constant. 

For $\gamma=5/3$, the generic equation of state (\ref{eq_state}) becomes~\cite{rund,anisinhom3,anisinhom4,anisinhom5} 
\begin{equation}\mu = mc^2\,n+\frac{3}{2}\,p,\qquad k\,T=\frac{p}{n} \label{eq_state_IG}\end{equation}
which is the equation of state of a non--relativistic limit of the classical ideal gas (Maxwell--Boltzmann gas). Another system that can be described by (\ref{eq_state}) is a suitable approximation to a mixture a non--relativistic and an ultra--relativistic gas~\cite{rund,fundam1,fundam2,anisinhom3,anisinhom4,anisinhom5}:
\ba \mu = m_{\rm{(nr)}}n_{\rm{(nr)}}c^2 +m_{\rm{(ur)}}c^2 n_{\rm{(ur)}}+\frac{3}{2}p_{\rm{(nr)}}+3p_{\rm{(ur)}}, \nonumber\\
 k\,T_{\rm{(nr)}}=\frac{p_{\rm{(nr)}}}{n_{\rm{(nr)}}},\quad k\,T_{\rm{(ur)}}=\frac{p_{\rm{(ur)}}}{n_{\rm{(ur)}}},\label{eq_state_mix}\nonumber\\\ea
where subindices ${}_{\rm{(nr)}}$ and ${}_{\rm{(ur)}}$ respectively stand for non--relativistic and ultra--relativistic. If we assume that $p_{\rm{(nr)}}\ll p_{\rm{(ur)}}$, but the non--relativistic gas provides the major contribution to rest mass ($ m_{\rm{(nr)}}n_{\rm{(nr)}}\gg  m_{\rm{(ur)}}n_{\rm{(ur)}}$), then (\ref{eq_state_mix}) becomes 
\begin{equation}\mu = m_{\rm{(nr)}}c^2\,n_{\rm{(nr)}}+3\,p_{\rm{(ur)}},\qquad k\,T_{\rm{(ur)}}=\frac{p_{\rm{(ur)}}}{n_{\rm{(ur)}}}, \label{eq_state_RG}\end{equation}
which is the equation of state (\ref{eq_state}) with $\gamma=4/3$. In practice, one uses (\ref{eq_state_RG}) to describe the so--called ``radiative gas'', which is a tightly coupled mixture of baryons and photons described as a single ``dust plus radiation'' fluid. In particular, since we neglect thermal motions of non--relativistic particles, $m_{\rm{(nr)}}c^2\,n_{\rm{(nr)}}$ could be the rest mass density of cold or ``warm'' dark matter and non--relativistic baryons, and so $m_{\rm{(nr)}}$ could be taken as the mass of a neutralino or another supersymmetric DM particle candidate.   

For either form (\ref{eq_state_IG}) or (\ref{eq_state_RG}), we will assume particle number conservation
\begin{equation}n^a = nu^a,\qquad \nabla_an^a=0,\qquad\Rightarrow\quad \dot n+n\,\Theta=0,\label{na_cons}\end{equation}
hence, if we consider a dark or warm DM gas described by (\ref{eq_state_IG}), we would be necessarily looking at dissipative effects after the ``freeze out'' era, when thermal equilibrium is no longer kept by particle annihilation~\cite{review,padma2,KT}. On the other hand, considering the radiative gas model, then (\ref{eq_state_RG}) with particle conservation is appropriate to describe the photon--electron interaction associated with Thomson or Compton scattering.
   
Since (\ref{Tab}) contains anisotropic pressure, which is not involved in the equation of state (\ref{eq_state}), it is natural to consider this pressure as a shear viscous stress associated to irreversible processes, whether in the classical ideal gas of WIMPS (\ref{eq_state_IG}) or in the radiative gas  (\ref{eq_state_RG}). Considering the fact that Extended Irreversible Thermodynamics (EIT) provides the most advanced theory complying with causality and stability~\cite{rund,JCVL,fundam1,fundam2}, we construct an entropy current $S^a$ within the framework of this theory. Since $\dot u_a=0$ for LTB spacetimes and the only dissipative stress is shear viscosity, the entropy current is
\begin{equation}S^a = S\,n^a=\left[S^{\rm{(eq)}}-\frac{c\,\tau\,\Pi_{ab}\Pi^{ab}}{2\eta\,n\,T}\right]\,n^a = \left[S^{\rm{(eq)}}-\frac{3\,c\,\tau\,\PP^2}{\eta\,n\,T}\right]\,n\, u^a\label{Sa}\end{equation}
where we used (\ref{PSEsc}), and $\tau,\,\eta$ are, respectively, the relaxation time and the coefficient of shear viscosity, while the specific entropy, $S^{\rm{(eq)}}$, is given by the equilibrium Gibbs equation
\begin{equation} T\dd S^{\rm{(eq)}}=\dd\left(\frac{\mu}{n}\right)+p\dd\left(\frac{1}{n}\right),\label{gibbs} \end{equation}
so its projection with respect to $u^a$ and the balance equation (\ref{ecbal1}) yield
\begin{equation} n\,T\,\dot S^{\rm{(eq)}}=-\sigma_{ab}\Pi^{ab}=-6\,\Sigma\,\PP.\label{gibbst} \end{equation}
The condition $\nabla_aS^a\geq 0$, together with (\ref{gibbst}), leads to the transport equation for shear viscosity~\cite{rund,JCVL,fundam1,fundam2}
\begin{equation} c\,\tau\,h_a^ch_b^d\dot\Pi_{cd}+\Pi_{ab}\left[1+\epsilon_0\eta\,T\,\tilde\nabla_c\left(\frac{c\,\tau}{2\,\eta\,T}\,u^c\right)\right]+2\,\eta\,\sigma_{ab}=0,\label{ectrans}\end{equation}
where $\epsilon_0=0,1$ is a ``switch'', so that (\ref{ectrans}) is the complete transport equation if $\epsilon_0=1$, and we get the ``truncated'' or Maxwell--Cattaneo equation if $\epsilon_0=0$. Using (\ref{Theta}) and (\ref{PSEsc}), equation (\ref{ectrans}) becomes the following scalar equation
\begin{equation} c\,\tau\,\dot\PP+\PP+2\,\eta\,\Sigma+\frac{\epsilon_0\,c\,\tau\,\PP}{2}\left[\frac{\dot\tau}{\tau}-\frac{\dot\eta}{\eta}-\frac{\dot T}{T}+\Theta\right]=0.\label{ectrans2}\end{equation}

To apply EIT to the non--relativistic and radiative gases, we need to substitute the equation of state (\ref{eq_state}) and utilize the forms of the coefficient of shear viscosity for these gases. From \cite{rund,JCVL,fundam1,fundam2}, we have 
\begin{equation}
\eta  = \alpha\, p\,c\,\tau ,\qquad \alpha  = \left\{ \begin{array}{l}
 1,\qquad \hbox{non--relativistic ideal gas} \\ 
 {\textstyle{4 \over 5}},\qquad \hbox{radiative gas} \\ 
 \end{array} \right.\label{eta}
\end{equation}
Hence, inserting $p=nkT$ and the particle conservation law (\ref{na_cons}), the transport equation (\ref{ectrans2}) becomes
\begin{equation} c\,\tau\,\left[\dot\PP+2\,\alpha\,p\,\Sigma+\epsilon_0\PP\frac{\dot T}{T}\right]+\PP=0,\label{ectrans3}\end{equation}
which clearly reveals how the difference between the complete and truncated equations can be dynamically significant, as it involves the term $\PP\,\dot T/T$. The entropy production subjected to the conservation law (\ref{na_cons}) follows readily from (\ref{Sa}) and (\ref{gibbst}) as
\begin{equation} \nabla_a S^a =n\dot S = 3\,k\,n\,\left[\frac{1}{c\,\tau}+(1-\epsilon_0)\frac{\dot p}{p}\right]\, \frac{\PP^2}{\alpha\,p^2},\label{ec_St3}\end{equation}
where we used (\ref{eta}) and (\ref{ectrans3}) to eliminate $\dot\PP$. 

In order to examine (\ref{Sa}) and (\ref{ectrans3}) we need to solve the field equations, or their equivalent ``fluid flow'' evolutions equations (\ref{ev_theta_13})--(\ref{cHam_13}), which would render the functional forms of the involved thermodynamical scalars. We look at this matter in the following section. 

\section{Quasi--local evolution equations.}

We can obtain an alternative set to the evolution equations (\ref{ev_theta_13})--(\ref{cHam_13}) that is completely equivalent, but easier to deal with numerically~\cite{sussQL}. This follows from using instead of the local scalars (\ref{loc_scals}), the scalar representation given by quasi--local variables $A_*$ defined by the map 
\begin{equation}\fl\JJ_*:X(\DD)\to X(\DD),\qquad A_*=\JJ_*(A)=\frac{\int_0^r{A R^2 R'\dd x}}{\int_0^r{R^2 R'\dd x}}.\label{QLmap}\end{equation}
where $X(\DD)$ is the set of all smooth integrable scalar functions $A$ defined in any spherical comoving region $\DD$ of the hypersurfaces $\T$ orthogonal to $u^a$, containing a symmetry center marked by $r=0$. The functions $A_*:\DD\to {\bf\rm{R}}$ that are images of $\JJ_*$ will be denoted by ``quasi--local'' (QL) scalars. In particular, we will call $A_*$ the QL dual of $A$. See \cite{sussQL} for a comprehensive discussion of of the map (\ref{QLmap}).  

Applying the map (\ref{QLmap}) to the scalars $\Theta$ and $\RR$ in (\ref{Theta}) and (\ref{RR}) we obtain their QL duals
\begin{equation} \Theta_* =\frac{3\dot R}{R},\qquad \RR_* =\frac{6K}{R^2}.\label{QL_TR}\end{equation}
Applying now (\ref{QLmap}) to $\mu$ and $p$, comparing with (\ref{mu1})--(\ref{p1}), and using (\ref{QL_TR}), these two field equations transform into 
\bse\ba \left(\frac{\Theta_*}{3}\right)^2 = \frac{\kappa}{3}\mu_* -\frac{\RR_*}{6},\label{QLfried}\\
\dot \Theta_*= -\frac{\Theta_*^2}{3}-\frac{\kappa}{2}\left(\mu_*+3p_*\right).\label{QLraych}\ea\ese
which are identical to the FLRW Friedman and Raychaudhuri equations, but among QL scalars. These equations can be further combined to yield identically the FLRW energy balance equation:
\begin{equation}\dot\mu_* = -\left(\mu_*+p_*\right)\,\Theta_*.\label{QLebal}\end{equation}
so that the QL scalars $\{\mu_*,\,p_*,\,\Theta_*\}$ effectively satisfy FLRW evolution laws. 

In order to relate local scalars to their and QL duals, we introduce the following ``relative deviations'' or ``perturbations''
\begin{equation} \Da \equiv \frac{A-A_*}{A_*},\quad \Rightarrow\quad A = A_*\,\left[1+\Da\right].\label{Da_def}\end{equation}
Therefore, all scalars $A$ in (\ref{loc_scals}) can be expressed in terms of their duals $A_*$ and perturbations $\Da$:
\ba \fl \mu = \mu_*\left[1+\Dm\right],\quad 
p = p_*\left[1+\Dp\right],\label{p_}\quad
\Theta = \Theta_*\left[1+\Dth\right],\quad
\RR = \RR_*\left[1+\DRR\right],\nonumber\\\fl\label{locvars_}\ea  
whereas $\Sigma,\,\PP$ and $\EE$ follow as
\bse\ba \Sigma &=& -\frac{1}{3}\,\left[\Theta-\Theta_*\right] = -\frac{1}{3}\,\Theta_*\,\Dth,\label{Sigma2}\\
\PP &=& \frac{1}{2}\,\left[p-p_*\right]= \frac{1}{2}\,p_*\,\Dp,\label{PP2}\\
\EE &=& -\frac{\kappa}{6}\,\left[\mu-\mu_* +\frac{3}{2}(p-p_*)\right]=-\frac{\kappa}{6}\,\left[\mu_*\Dm +\frac{3}{2}p_*\Dp\right],\label{EE2}\ea\ese
which leads to an alternative QL scalar representation $\{A_*,\,\Da\}$ that it is fully equivalent to the local representation. We derive now the evolution and constraint equations for this representation.  

From differentiating both sides of (\ref{QLmap}) and using (\ref{Da_def}), we can relate radial gradients of $\mu_*,\,p_*$ and $\HH_*$ with their corresponding $\delta$ functions by
\begin{equation} \frac{\Theta_*{}'}{\Theta_*} = \frac{3R'}{R}\,\Dth,\qquad \frac{\mu_*{}'}{\mu_*} = \frac{3R'}{R}\Dm,\qquad \frac{p_*{}'}{p_*} = \frac{3R'}{R}\Dp, \label{rad_grads}\end{equation}
while (\ref{QLraych}) and (\ref{QLebal}) are evolution equations for $\dot\mu_*$ and $\dot\Theta_*$. Hence, the evolution equations for $\Dm$ and $\Dth$ follow from the consistency condition $[A_*']\,\dot{}=[\dot A_*]'$ applied to (\ref{QLraych}), (\ref{QLebal}) and (\ref{rad_grads}) for $A=\Theta_*$ and $\mu_*$. The result is the following set of autonomous evolution equations for the QL scalar representation $\{A_*,\,\Da\}$:
\bse\ba 
\fl \dot\mu_* &=& -\left[\,1+w\,\right]\,\mu_*\,\Theta_*,\label{evmu_ql}\\
\fl \dot\Theta_* &=& -\frac{\Theta_*^2}{3} -\frac{\kappa}{2}\,\left[\,1+3\,w\,\right]\,\mu_*,
\label{evHH_ql}\\
\fl \dDm &=& \Theta_*\,\left[\left(\Dm-\Dp\right)\,w-\left(1+w+\Dm\right)\Dth\right],
\label{evDmu_ql}\\
\fl \dDth &=& -\frac{\Theta_*}{3}\,\left(1+\Dth\right)\,\Dth + \frac{\kappa\mu_*}{6\,(\Theta_*/3)}\left[\Dth-\Dm+3w\,\left(\Dth-\Dp\right)\right],
\label{evDth_ql}
\ea\ese
where $w\equiv p_*/\mu_*$. 
 
The constraints associated with these evolution equations are simply the spatial  gradients (\ref{rad_grads}), while the Friedman equation (or Hamiltonian constraint) is (\ref{QLfried}). Notice that (\ref{rad_grads}) follow directly from differentiating the integral definition (\ref{QLmap}), so by using the QL variables we do not need to solve these constraints in order to integrate (\ref{evmu_ql})--(\ref{evDth_ql}).  

It is straightforwards to  prove that the evolution equations (\ref{evmu_ql})--(\ref{evDth_ql}) and the constraints (\ref{QLfried}) and (\ref{rad_grads}) are wholly equivalent to the evolution equations (\ref{ev_theta_13})--(\ref{cHam_13}) of the fluid flow formalism of Ellis, Bruni and Dunsbury \cite{1plus3}.  It is also important to mention that the QL representation $\{A_*,\,\Da\}$ leads to a characterization of LTB spacetimes as exact, non--linear, gauge invariant and covariant perturbations over a FLRW formal background defined by the QL scalars $A_*$, which satisfy FLRW dynamics. See ~\cite{sussQL} for details. 

\section{Evolution equations for the generic ideal gas.}

In order to integrate the system (\ref{evmu_ql})--(\ref{evDth_ql}) we need to prescribe a relation between $\mu_*,\,p_*$ and $\Dm,\,\Dp$. Since we are interested in thermal dissipative phenomena characteristic of a hydrodynamical regime of short range interactions, the physically meaning full equation of state (\ref{eq_state}) is that relating local variables $\mu$ and $p$, and not QL variables. However, (\ref{eq_state}) is a linear functional relation, hence its validity as a local relation and the assumption of particle current conservation (\ref{na_cons}) are sufficient conditions to render (\ref{evmu_ql})--(\ref{evDth_ql}) a fully determined system in which the QL variables are basically auxiliary variables (and the physical variables are the local ones).

Assuming the local equation of state (\ref{eq_state}) and using (\ref{Da_def}) with $A=\mu,\,n,\,p$ leads to the following conditions on the QL variables
\bse\ba \mu_* &=& m\,c^2\,n_*+\frac{p_*}{\gamma-1},\label{eq_state_ql1}\\
\Dm &=& \frac{m\,c^2\,n_*}{\mu_*}\,\Dn+\frac{p_*}{(\gamma-1)\,\mu_*}\,\Dp,\label{eq_state_ql2}\ea\ese   
Using the particle numbers conservation law (\ref{na_cons}) with $n=n_*(1+\Dn)$, together with (\ref{eq_state_ql1})--(\ref{eq_state_ql2}), transforms (\ref{evmu_ql})--(\ref{evDth_ql}) into the fully determined system
\bse\ba \fl \dot n_* &=& -n_*\,\Theta_*,\label{evn_ql}\\
        \fl \dot p_* &=& -\gamma\,\,p_*\,\Theta_*,\label{evp_ql}\\
       \fl \dot \Theta_* &=& -\frac{\Theta_*^2}{3}-\frac{\kappa}{2}\,\left[mc^2\,n_*+\gamma_1\,p_*\right],\label{evZ_ql}\\
       \fl \dDn &=& -\left(1+\Dn\right)\,\Theta_*\,\Dth,\label{evDn_ql}\\
       \fl \dDp &=& -\left(\gamma+\Dp\right)\,\Theta_*\,\Dth,\label{evDp_ql}\\
      \fl  \dDth &=& -\frac{1}{3}\left(1+\Dth\right)\,\Theta_*\,\Dth-\frac{\kappa}{6}\,\left[mc^2\,n_*\left(\Dn-\Dth\right)+\gamma_1\,p_*\left(\Dp-\Dth\right)\right],\nonumber\\\fl\label{evdZ_ql}\ea\ese
with $\gamma_1\equiv (3\gamma-2)/(\gamma-1)$.
Once the system (\ref{evn_ql})--(\ref{evdZ_ql}) is solved numerically for appropriate initial conditions (see appendices of \cite{sussQL}), we obtain the local variables $p,\,\Theta,\,\Sigma,\,\PP$ from (\ref{locvars_}), (\ref{Sigma2}) and (\ref{PP2}), while the temperature $T$ follows readily as 
\begin{equation}  k\,T = \frac{p}{n}=\frac{p_*\,\left[1+\Dp\right]}{n_*\,\left[1+\Dn\right]},\label{kTql}\end{equation}
With the help from (\ref{locvars_}), (\ref{Sigma2}), (\ref{PP2}), (\ref{evn_ql})--(\ref{evdZ_ql}) and (\ref{kTql}), the transport equation (\ref{ectrans3}) reduces to the following two algebraic constraints defining the relaxation times for the full ($\epsilon=1$) and truncated ($\epsilon=0$) cases:
\ba c\tau &=& \frac{3\Dp\left(1+\Dp\right)}{\Theta_*\Dth\,\left[4\alpha(\Dp)^2+(3+8\alpha)\Dp+3\gamma+4\alpha\right]},\qquad \epsilon=1,\label{tau1}\\
c\tau &=& \frac{3\Dp}{\Theta_*\,\left[((4\alpha+3)\Dth+3\gamma)\Dp+(4\alpha+3\gamma)\Dth\right]},\qquad \epsilon=0,\label{tau2}\ea
while the entropy production law (\ref{ec_St3}) leads to
\begin{equation} \fl \dot S = \frac{3k(\Dp)^2}{4\alpha\,[1+\Dp]^2}\,\left[\frac{1}{c\tau}+\frac{(\epsilon_0-1)\Theta_*[(\gamma+\Dth)\Dp+(1+\Dth)\gamma}{1+\Dp}\right].\label{ec_St4} \end{equation}
However, substituting $\tau$ from either (\ref{tau1}) or (\ref{tau2}) into (\ref{ec_St4}) we obtain the same expression of $\dot S$ for the full and truncated cases:
\begin{equation} \dot S = \frac{k\,\Theta_*\,\Dth\,\Dp\,[\,4\alpha\,(\Dp)^2+(3+8\alpha)\,\Dp+4\alpha+3\gamma]}{4\alpha\,[1+\Dp]^3}. \label{ec_St5}\end{equation}
Dissipative effects associated with shear viscosity for thermal systems associated with (\ref{eq_state}) can be now examined by using the numerical solution of (\ref{evn_ql})--(\ref{evdZ_ql}) to calculate the relaxation time scale given by (\ref{tau1}) or (\ref{tau2}), as well as the entropy production $n\dot S$ from (\ref{ec_St4}). 
\begin{figure}[htbp]
\includegraphics[width=4in]{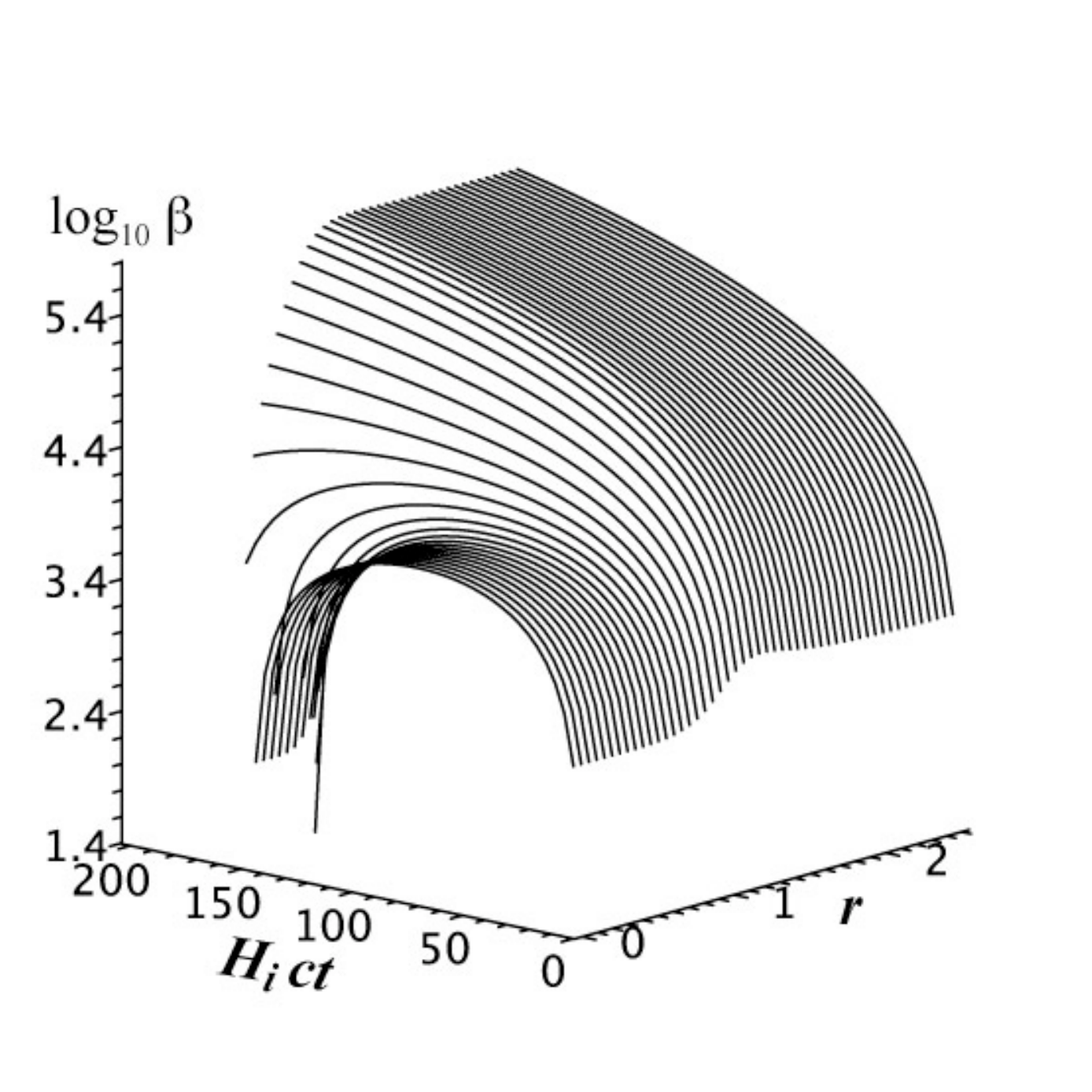}
\caption{{\bf Coldness parameter $\beta$.} The figure displays the function $\beta=mc^2/(kT)$ given by (\ref{beta}) for the ideal gas of WIMPS configuration described in section \ref{numres}. The layers near the center ($r=0$) bounce and collapse to a black hole where thermal motions dominate rest mass ($\beta\to 0$), though the hydrodynamical regime is no longer valid in this stage. }
\label{fig1}
\end{figure}
\begin{figure}[htbp]
\includegraphics[width=4.5in]{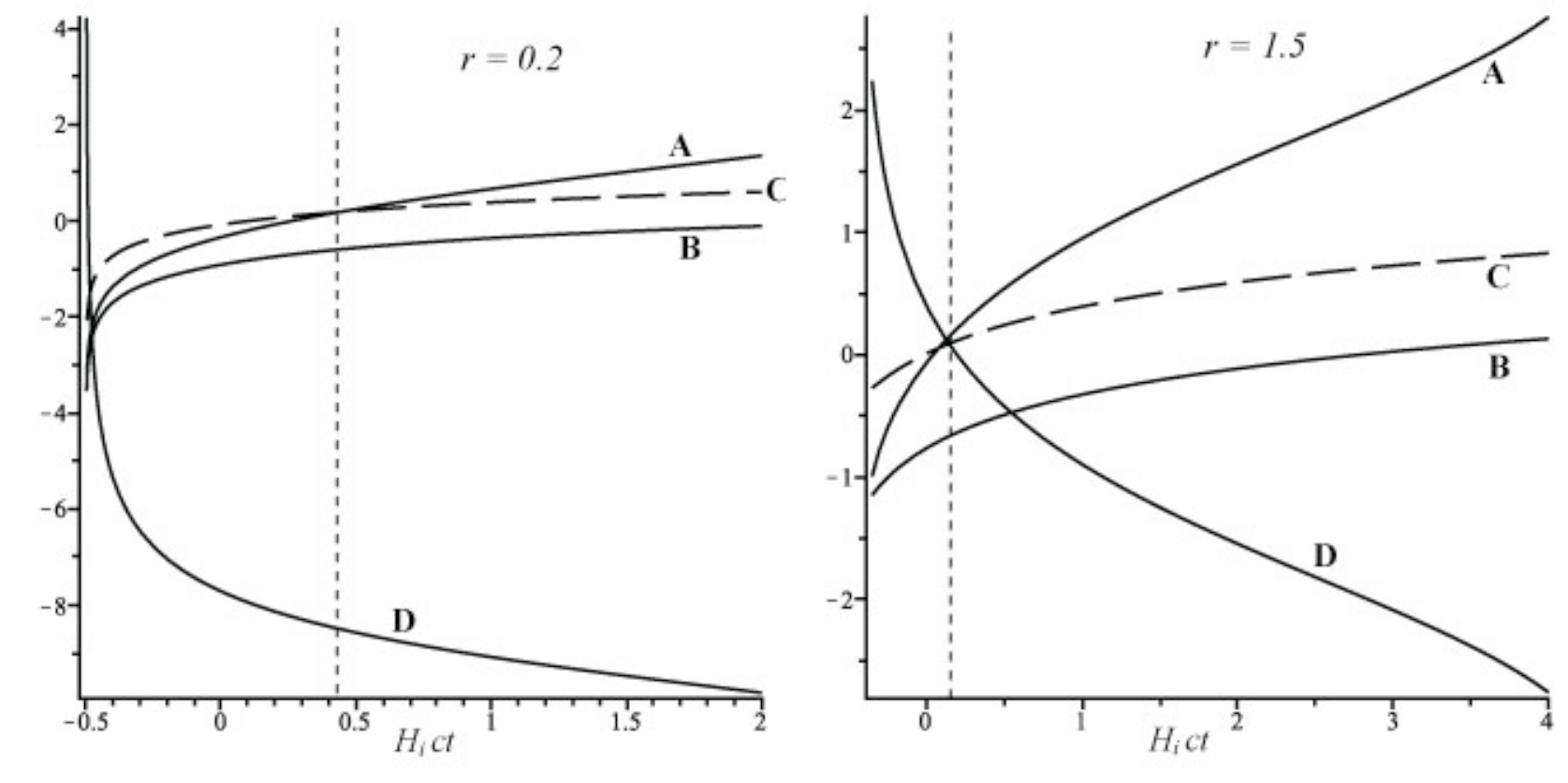}
\caption{{\bf Relaxation vs Hubble times.} The figure depicts the logarithm of $\tau$ for the full transport equation (A), for the truncated version (B), the Hubble time $3/\Theta$ (C) and $\dot S$ (D). The vertical dotted line depicts the extension of the relaxation time scale up to $\tau\sim 3/\Theta$. Notice how for central (over--density) layers (left panel), with more thermal energy (lesser $\beta$), this time scale has a much larger extent than in the layers corresponding to the cosmic background (right panel). }
\label{fig3}
\end{figure}
\begin{figure}[htbp]
\includegraphics[width=4.5in]{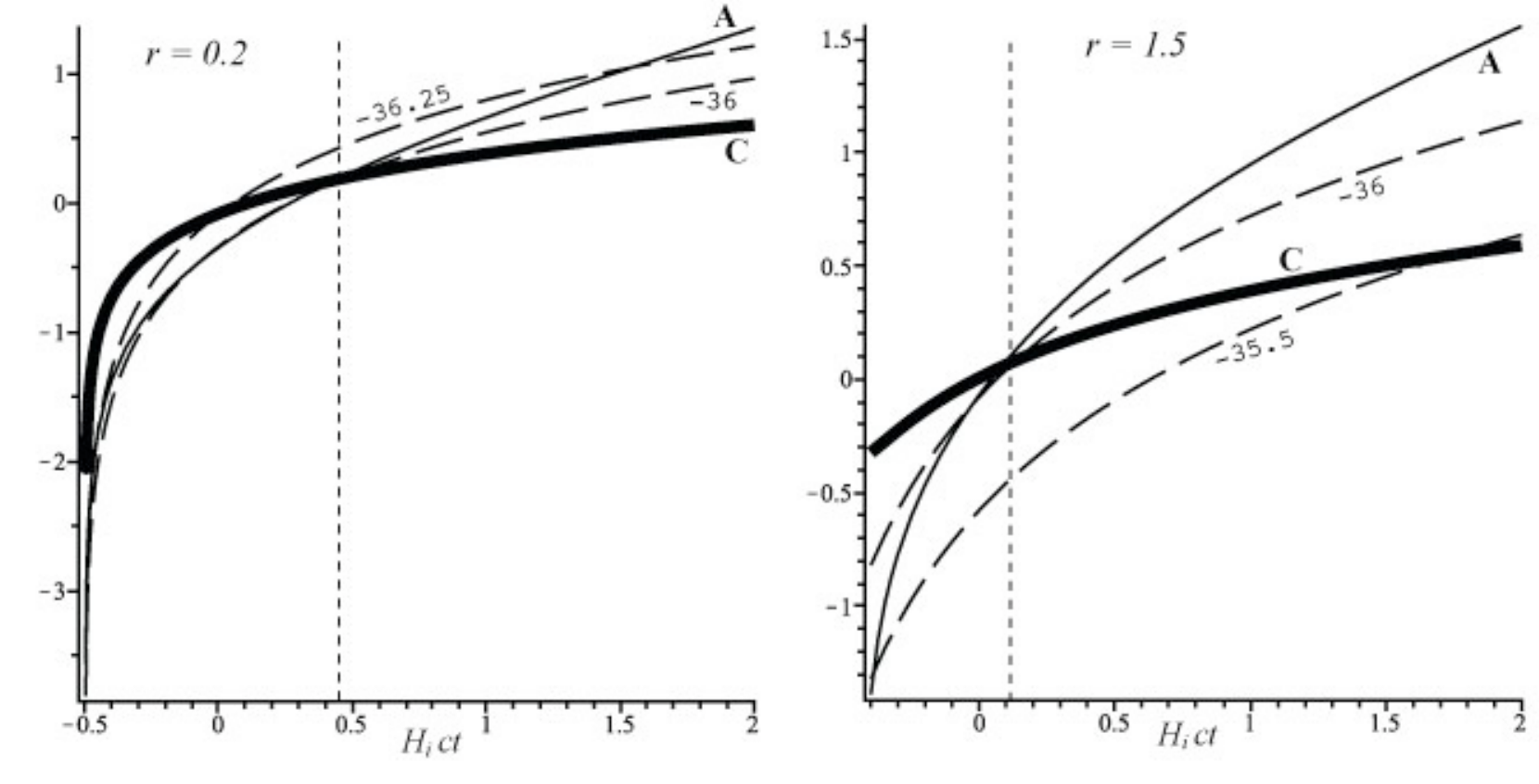}
\caption{{\bf Relaxation vs collision times.} The figure depicts the logarithm of $\tau$ for the full transport equation (A), the Hubble time $3/\Theta$ (C) and collision times (dotted curves) given by (\ref{sigma}) with the numbers indicating the exponent $s_0$, and with the WIMP mass--energy of 100 GeV. Notice how $\tau$ is qualitatively analogous to collision times with cross section areas $\sigma\sim 10^{-36} \hbox{cm}^2$ consistent with weak interactions of cold dark matter WIMPS. As in figure 2, the relaxation time scale has a larger extent in layers in the over--density (left panel) than in the cosmic background (right panel). }
\label{fig3}
\end{figure}
\section{The gas of WIMPS} 

Dissipative phenomena associated with shear viscosity in spacetimes with LTB metrics have been studied mostly on the radiative gas model \cite{anisinhom4,anisinhom5} (but see \cite{anisinhom3}). In particular, the comparison between relaxation and collision times was examined in \cite{anisinhom5} for this model in the context of the cosmological radiative era.  In this article we consider the same issue, but for a gas of non--relativistic cold dark matter particles (WIMPS) after its decoupling or ``freeze out'' from the cosmic fluid, when thermal equilibrium is no longer maintained by particle annihilation~\cite{review,padma2,KT}. Since cold dark matter has no effect on cosmic nucleosynthesis, this decoupling must have happened before nucleosynthesis at around $t\sim 200$ sec, and so the gas of WIMPS can be described as an ideal gas in the non--relativistic limit, corresponding to the equation of state (\ref{eq_state_IG}). Hence, the expressions for the coefficient of shear viscosity, relaxation times and entropy production are (\ref{eta}), (\ref{eq_state_ql1})--(\ref{ec_St5}), for the the values $\gamma=5/3$ and $\alpha=1$. 

The relaxation time is a mesoscopic quantity that could be, in principle, obtained by means of collision integrals~\cite{JCVL}, but cannot be given in terms of an ``equation of state'' relating macroscopic thermodynamical scalars. Usually, this quantity is taken simply as a mean collision time, or it is assumed to have the same order of magnitude value as these times. However, as shown by the results of \cite{anisinhom5} in the radiative gas model, there is no reason for this to be the case. Since these two time scales follow from physically distinct concepts, they must be different functions that could exhibit qualitatively analogous behavior and/or could be of the same order of magnitude. 

For an ideal gas the mean collision time is given as~\cite{rund,padma2,KT}
\begin{equation} ct_{\rm{col}}=\frac{1}{\sigma\,n}=\frac{1}{\sigma\,n_*\,(1+\Dn)},\label{tcol} \end{equation}
where $\sigma=\sigma(n,T)$ is the collision cross section area, whose precise functional form follows from the specific particle interactions involved in the gas model. For a gas of WIMPS, we can identify a decoupling stage as cosmic times $ct=ct_{\rm{D}}$ for which the reaction times compare with the Hubble expansion time $t_{H}$
\begin{equation} n\,\sigma(n,T) \approx c\,t_{H}\sim \frac{3}{\Theta},\label{tD}\end{equation}
so that for $t<t_{\rm{D}}$, before its decoupling from the cosmic fluid, $\sigma$ is associated with particle pair annihilations and its form follows from theoretical considerations pertinent to supersymmetric cold dark matter candidate particles~\cite{padma2,KT}. Moreover, we will examine dissipative effects in the gas of WIMPS for $t>t_{\rm{D}}$, after this freeze out when particle numbers are conserved. The justification for these after freeze out dissipative processes comes from the assumption that there could have been dissipation in the earlier stage $t<t_{\rm{D}}$, and so it is reasonable to assume that once particle annihilations stop at (\ref{tD}), there should be a short timed relaxation process characterized by a weak self--interaction associated with very small cross section areas, so that after this process the fluid becomes completely non--collisional. Since this relaxation should be of short duration, we can model these cross section areas empirically by the simple ansatz~\cite{KT,padma2}
\begin{equation}  \sigma \sim 10^{s_0}\,\hbox{cm}^2,\qquad -40 < s_0 < -34.\label{sigma}\end{equation}
Hence, we expect $\tau$ in (\ref{tau1}) and (\ref{tau2}) to exhibit a qualitatively analogous behavior as (\ref{tcol}) for cross section areas having magnitudes given by (\ref{sigma}). In particular, the existence of an interaction that can be associated with shear viscosity requires that these time scales are of lesser magnitude than the Hubble time: $c\tau < 3/\Theta$ and $ ct_{\rm{col}} < 3/\Theta$, with the relaxation time scale given by the cosmic time $ct$ such that  $c\tau \sim 3/\Theta$ and $ct_{\rm{col}}\sim 3/\Theta$, and thus,  $c\,\tau\sim ct_{\rm{col}}$ at these cosmic time values. Hence, when $c\tau > 3/\Theta$ the gas expands in a non--collisional stage. However, for earlier times $c\tau$ and $ ct_{\rm{col}}$ need not be the same function, just have comparable magnitudes. Also, for the relaxation time scale in which $c\tau < 3/\Theta$, we must have necessarily $\dot S>0$, so that there is entropy production with $\dot S\to 0$ as $c\tau$ and $ct_{\rm{col}}$ overtake $3/\Theta$ and entropy becomes a maximum associated with equilibrium conditions. 

In order to test numerically these conditions, we define the following dimensionless variables associated with $n_*,\,p_*$ and $\Theta_*$ in (\ref{evn_ql})--(\ref{evdZ_ql}) (notice that the $\delta$ functions are already dimensionless):
\begin{equation} x \equiv \frac{\kappa m\,c^2\,n_*}{3\,H_i^2},\qquad y\equiv \frac{\kappa p_*}{3H_i^2},\qquad z\equiv \frac{\Theta}{3\,H_i},\label{dimless}\end{equation}
where $H_i\sim 1/(c t_i)$ is taken as the Hubble scale factor for the initial time surface $t_i\sim 200$ sec., and we will consider $m c^2 = 100$ GeV to be the rest mass--energy of the WIMP. In terms of (\ref{dimless}), the collision time (\ref{tcol}) is given by
\begin{equation} \fl ct_{\rm{col}} = \frac{\kappa m c^2}{\sigma\,x\,(1+\Dn)}= 4.3\times 10^{-26}\times \frac{m\,c^2}{\hbox{GeV}}\times \frac{\hbox{cm}^2}{10^{s_0}}\times \frac{1}{x\,(1+\Dn)}.\end{equation}
We will examine in the following section these different time scales associated with the relaxation scale using the numeric solutions of (\ref{evn_ql})--(\ref{evdZ_ql}).

\section{Relaxation time scales: numeric results.}\label{numres}

In order to set up appropriate initial conditions for $x,\,y$ and $z$, we use equations (\ref{QLfried}) and (\ref{eq_state_ql1}) for $\gamma=5/3$, leading to
\begin{equation} z_i^2(r)=x_i(r)+y_i(r)-k_i(r),\qquad k_i(r)=\frac{[\RR_*]_i}{6\,H_i^2},\label{init_fried}\end{equation}
where the subindex ${}_i$ denotes evaluation at $t=t_i$. Initial conditions for a central over--density with small positive curvature that smoothy blends to a cosmic background with small negative spatial curvature can be achieved by choosing $k_i(r)$ as any smooth function for which $k_i(0)=0.1$ and  $k_i(r)\to -0.1$ for $r\to\infty$. Central and asymptotic values for $x_i$ and $y_i$ are given by
\begin{equation}\fl  x_i(0) = 1.5,\qquad x_i(\infty) = 0.9,\qquad\qquad
y_i(0) = 0.08,\qquad y_i(\infty) = 0.02.\label{init_xy}\end{equation}
The form of $z_i$ follows from (\ref{init_fried}) and (\ref{init_xy}), while the forms for the initial value functions $[\Dn]_i,\,[\Dp]_i$ and $[\Dth]_i$ can be obtained from $x_i,\,y_i,\,z_i$ by means of (\ref{rad_grads}) evaluated at $t=t_i$ (see the appendices of \cite{sussQL}). 

An important parameter in thermal systems is the ``coldness'' parameter
\begin{equation} \beta = \frac{m\,c^2}{k\,T}=\frac{x\,[1+\Dn]}{y\,[1+\Dp]},\label{beta}\end{equation}
which, with the numeric values of (\ref{init_xy}), takes initial values $\beta_i(0)\sim 20$ and $\beta_i(\infty)\sim 40$, which are reasonable values for cold dark matter WIMPS that are non--relativistic when they decouple at $t=t_{\rm{D}}$~\cite{review,KT,padma2}.  We display in figure 1 the function $\beta$ that results from the numeric solution of (\ref{evn_ql})--(\ref{evdZ_ql}) for the configuration outlined above. As the configuration expands we can see how $\beta$ increases for all $r$ to clear non--relativistic values  $\beta\gg 1$, but layers in the over-density region (around $r=0$) collapse to a black hole at around $H_i ct\sim 150$, with $\beta\to 0$, indicating dominance of internal energy density over rest mass energy density near the final collapse. However, the hydrodynamical regime is no longer a valid approximation in this stage, as WIMP configurations do not evolve to black holes. In more realistic structure formation scenarios the WIMP gas becomes non--collisional and undergoes non--collisional relaxation phenomena, such as virialization~\cite{padma2}, leading to stable bound structures.

The relaxation of a viscous dissipative stress requires that $\dot S>0$ while  $c\,\tau < 3/\Theta$, but both $\tau$ and $t_{\rm{col}}$ must overtake $3/\Theta$ as $\dot S\to 0$. We test these conditions numerically in figure 2, for two different values of $r$ (at the over--density in the left panel and at the cosmic background in the right panel), and for the relaxation times of the full (\ref{tau1}) and truncated (\ref{tau2}) transport equations and for $\dot S$ given by (\ref{ec_St5}). As shown by this figure, we have $c\,\tau < 3/\Theta$ for all times for the relaxation time (\ref{tau2}) of the truncated equation. Therefore, the relaxation time for the truncated (Maxwell--Cattaneo) does not exhibit the appropriate behavior of a relaxation parameter, which means that the full transport equation is needed to provide an adequate description of the transient dissipative phenomena for the ideal gas of WIMPS. The same result was obtained for the radiative gas model in \cite{anisinhom5}. On the other hand, the relaxation time (\ref{tau1}) of the full transport equation exhibits the expected behavior and overtakes the Hubble time $3/\Theta$ as $\dot S\to 0$. We show in figure 3 how the relaxation time (\ref{tau1}) of the full transport equation (for $mc^2=100$ GeV) in the whole relaxation time scale is qualitatively analogous to collision times with cross sections given by (\ref{sigma}) with $s_0\sim -36$, which characterize expected weak interactions for decoupled WIMPS~\cite{review,padma2,KT}. 
      
\section{Conclusion}

We have examined causal dissipation from shear viscosity in the context of a large class of inhomogeneous spherically symmetric spacetimes described by the LTB metric (\ref{ltb}). A generic equation of state was suggested, which contains as particular cases the classical, non--relativistic, ideal gas, as well as the radiative gas in the approximation in which thermal motions of the non--relativistic species are ignored. We obtained a set of evolution equations  equivalent to the field and balance equations, whose numeric solutions can be used to compute the relaxation times for the full and truncated transport equations, the rate of change of specific entropy and collision times for suitable cross section areas. We considered the non--relativistic ideal gas as an appropriate equation of state for a gas of cold dark matter WIMPS undergoing a transition to equilibrium soon after their freeze out and decoupling from the cosmic fluid at the outset of cosmic nucleosynthesis. The comparison between relaxation and collision times yielded similar results as those obtained in \cite{anisinhom5} with the radiative gas model, namely, that only the relaxation time from the full transport equation exhibits the expected behavior of a relaxation parameter, being also qualitatively analogous and of the same order of magnitude as collision times with reasonable cross sections for a gas of WIMPS. This result is shown in figures 2 and 3. 

It is evident that the study of shear viscosity without other dissipative fluxes (heat flux and bulk viscosity) is an idealized situation which follows from the constraints of the LTB metric. Although the inhomogeneous conditions provided by this metric are mathematically tractable, they are not trivial and contain enough structure to examine non--linear effects that cannot be studied in a FLRW context or with linear perturbation. Another shortcoming is the use the transport equation itself to define the relaxation times, as it was done in \cite{anisinhom3,anisinhom4,anisinhom5}, instead of using it as a free parameter to be specified. The resulting expressions (\ref{tau1}) and (\ref{tau2}) are, evidently, approximations to the actual relaxation times, but this approximation will be reasonable if the obtained quantities behave as a relaxation parameters. As shown in section 7 and in figures 2 and 3, the relaxation time for the full equation does  exhibit the expected behavior, and so this approximation is reasonable. Future work along these lines would necessarily require a more general metric framework and more elaborated numerical methods. This work is presently under consideration.

\end{document}